\documentclass[aps,prl,twocolumn,groupedaddress]{revtex4}

\usepackage{graphicx}
\usepackage{bm}

\begin{document}

\title{An atom mirror etched from a hard drive}
\author{Benjamin Lev}
\email[Electronic mail:  ]{benlev@caltech.edu}
\author{Yves Lassailly}
\altaffiliation[Present address:  ]{Laboratoire PHMC / Ecole
Polytechnique, 91128 Palaiseau cedex, France}
\author{Chungsok Lee}
\author{Axel Scherer}
\author{Hideo Mabuchi}
\affiliation{California Institute of Technology, Pasadena, CA}

\date{\today}

\begin{abstract}
We describe the fabrication of an atom mirror by etching of a
common hard drive, and we report the observation of specular
retroreflection of 11 $\mu$K cesium atoms using this mirror. The
atoms were trapped and cooled above the hard drive using the
mirror magneto-optical trap technique, and upon release, two full
bounces were detected. The hard drive atom mirror will be a useful
tool for both atom optics and quantum computation.
\end{abstract}

\pacs{don't know}

\maketitle

Laser cooling and trapping techniques have made possible the
preparation of extremely cold samples of atoms~\cite{Metcalf99}.
Atom optics employs elements such as mirrors, lenses, gratings and
beam-splitters to manipulate these cold atoms in a fashion similar
to the familiar photon optics. The advent of the atom
laser~\cite{Ketterle} has enhanced the importance of developing
atom optical elements. In particular, atom mirrors---surfaces that
reflect atoms---play a crucial role in the field of atom optics,
and it is of keen interest to develop mirrors that are simple to
fabricate yet highly specular. In this Letter we demonstrate a
straightforward technique to produce large area, high resolution
permanent-magnetic structures on flat, rigid, and inexpensive
substrates.

Several types of atom mirrors have been fabricated using
evanescent light fields~\cite{Aspect02}, dynamic magnetic
fields~\cite{Boshier}, and static magnetic fields~\cite{Hindsrev}.
Evanescent mirrors repulse atoms from a prism surface using a
potential created by a blue-detuned light field. Although magnetic
mirrors cannot generally be modulated as easily as evanescent
mirrors, they do offer many advantages: passive operation,
compactness (no laser access is needed), and much larger repulsive
areas.

Magnetic mirrors employ a sheet of alternating current or magnetization to
create an exponentially increasing potential near the mirror
surface~\cite{Opat}. To lowest order, this potential is proportional to $B_0
e^{-ky}$.  The surface field, $B_0$, sets the maximum atom energy that can be
reflected, and the spatial period of the current or magnetization, $a=2\pi/k$,
determines the amount of time the atoms interact with the mirror.  The
magnetic mirror approximates a perfectly flat mirror as $B_0$ increases and
$a$ decreases.  For example, if $B_0=1$ kG and $a=1$ $\mu$m, a cesium atom in
the $6^2S_{1/2}$ $F=4$, $m_F=4$ state will be reflected when dropped from a
height of 0.4 m, and will only interact with the mirror for 5 $\mu$s if
dropped from 2 cm.

Mirrors made from serpentine patterns of wires can produce
time-dependent reflection potentials. However, they have not been
fabricated with periods smaller than 10 $\mu$m, and the power
dissipated by the small wires requires cooling by liquid nitrogen
and pulsed operation~\cite{Westervelt98}. Sinusoidal magnetization
of audio-tape, floppy disks, and videotape can produce magnetic
mirrors with magnetization periods down to 12
$\mu$m~\cite{Hindsrev}. Mirrors made from millimeter-sized arrays
of permanent magnets have been demonstrated, as have mirrors
produced by 1 to 4 $\mu$m periodic structures fabricated by
sputtering ferromagnetic material onto a grooved substrate
patterned by electron-beam lithography~\cite{Hannaford02}.

We recently fabricated a magnetic mirror by etching a common hard
drive, and we have used this mirror to retroreflect a cold cloud
of $10^6$ cesium atoms. Hard drives offer several advantages for
making and using atom mirrors.  The common hard drive provides a
large surface area of thin magnetic film whose surface is
specifically designed to be very flat, smooth, and rigid.
Furthermore, the film's remnant magnetic field and coercivity can
be as large as 7 kG and 3 kG, respectively~\cite{Comstock}. An
atom mirror could in principle be fabricated with a 2 $\mu$m
periodicity over the entire surface of the hard drive. Old or
discarded hard drives may be used: an Apple hard drive from the
mid-1990's was used for the experiment presented here.

\begin{figure}
\scalebox{.6}[.6]{\includegraphics{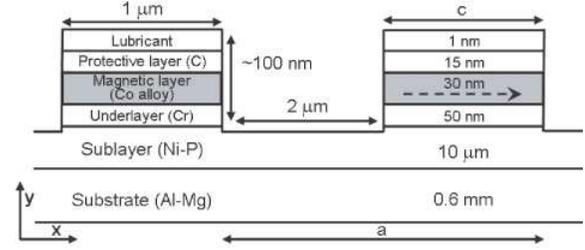}}
\caption{\label{fig:crossection}A cross-section of the etched hard
drive.  The magnetization is in plane.  See Ref.~\cite{Comstock}
for a description of the hard drive layers.}
\end{figure}

We fabricate the mirror by etching 2 $\mu$m wide, $\sim100$ nm
deep trenches into a 1 cm$^2$ section of the surface of the hard
drive. These 100 nm trenches extend past the magnetic layer to
form a periodic array of 1 $\mu$m wide, 30 nm thick, and 1 cm long
stripes of cobalt alloy (see Fig.~\ref{fig:crossection}). The
cobalt alloy is granular, which enhances the coercivity and allows
us to magnetize the material in plane and parallel to the short
axis of the magnetic strips.  We do not know the exact materials
and thicknesses of the layers of the proprietary hard drive.
However, etching $\sim100$ nm is sufficient to remove the magnetic
layer.

Standard photolithography is used to create the etch mask. After
cutting the hard drive into 2 to 3 cm$^2$ sections, positive
photoresist (TSMR-8900 from Tokyo Ohka Kogyo Co.) is spun onto the
cleaned hard drive surface for 40 s at 4200 rpm.  A 15 s exposure
followed by 65 s in the developer (NMD-W 2.38\%) maps the
photomask lines into the resist.  The sample is ion milled with
argon in a inductively coupled plasma (ICP) system. We etch for 8
minutes at a forward power of 100 W, ICP power of 400 W, and an
argon flow of 40 sccm. The remaining photoresist is removed with
acetone and, if necessary, a soft swab.  To erase the hard drive's
bits and magnetize it as a mirror, we insert the hard drive
section into the field of an 8 kG electromagnet whose field is
parallel to the surface and perpendicular to the magnetic stripes.

The magnetic field from the etched hard drive, with in-plane magnetization,
$M_0$, parallel to the short axis of the magnetic stipes, is analogous to a
periodic sheet of alternating in-plane magnetization $+M_0/2$ and $-M_0/2$. In
the infinite array limit, the magnetic field above the surface is
\begin{equation}
B^2=B_1^2e^{-2ky}+2B_1B_3\mbox{cos}(2kx)e^{-4ky}+B_3^2e^{-9ky}+...,
\end{equation}
where $B_1=\mu_0 M_0(1-e^{-kb})/\pi$, $B_3=\mu_0
M_0(1-e^{-3kb})/3\pi$, and $b=$ 30 nm is the thickness of the
magnetic layer.  The field has no components in the z-direction,
and rotates with a period equal to $a$ in the x-y plane.  Cesium
atoms in the $F=4$, $m_F=4$ state, which has the largest
weak-field seeking magnetic moment, would have to be dropped from
a height of 25 cm to penetrate to a height at which the second
term in the expansion is equal the first, so to a good
approximation the field may be written as
\begin{equation}\label{onetoonemag}
B\approx B_1 e^{-ky} + B_3 e^{-3ky}\mbox{cos}(2kx).
\end{equation}
For our hard drive mirror, $B_1$ is equal to 2 to 4 kG depending on the
specific cobalt alloy.  When $a=1$ $\mu$m, the ratio of the first harmonic
term to the purely exponential term for a cesium atom dropped from a height of
2 mm (20 mm) is $1\times10^{-6}$ ($1\times10^{-3}$) at the turning point
$y=0.8$ $\mu$m ($y=0.4$ $\mu$m).

\begin{figure}
\scalebox{1}[1]{\includegraphics{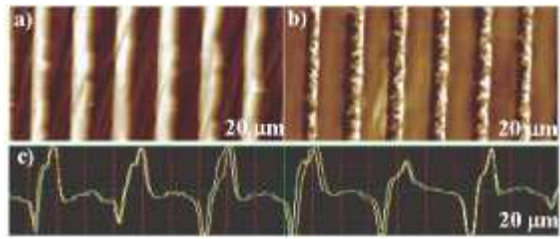}}
\caption{\label{fig:AFMMFM} Twenty micron wide (a) AFM scan, (b)
MFM scan, and (c) MFM cross-section of the etched hard drive
surface.}
\end{figure}

The etched hard drive used for the experiment has $a\approx3$
$\mu$m and $c\approx1$ $\mu$m resulting in a ratio of magnetic
layer to gap that is approximately 1:2.
Figures~\ref{fig:AFMMFM}(a) and (b) show 20 $\mu$m wide AFM and
MFM scans of the hard drive surface. The trenches in the AFM scan
are dark, and the light to dark variation of magnetic strips shows
the north and south poles of the magnetization.
Figure~\ref{fig:AFMMFM}(c) shows a 20 $\mu$m cross-section of the
MFM scan:  peaks represent the north and south poles. To describe
the field above our etched hard drive, Eqn.~\ref{onetoonemag} can
be modified to account for the deviation from a 1:1 width ratio by
multiplying $B_1$ by $\mbox{sin}(\pi c/a)$ and $B_3$ by
$\mbox{sin}(3\pi c/a)$.  In our device, the ratio of
$c/a\approx1/3$ decreases the $B_1$ term by 0.9, but causes the
corrugation term to nearly vanish.

\begin{figure}
\scalebox{.85}[.85]{\includegraphics{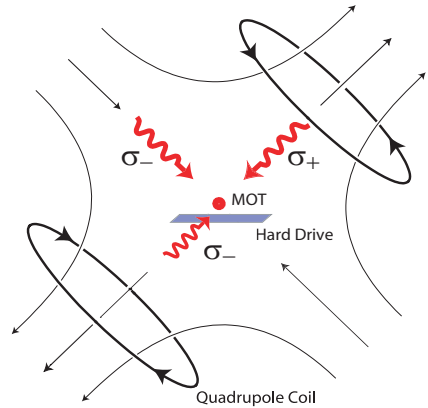}}
\caption{\label{fig:mMOT} Diagram of the experimental set-up.  A
quadrupole field and two 45$^\circ$ laser beams and one
retroreflected grazing beam form a mirror MOT 1.5 to 4 mm above
the etched hard drive.}
\end{figure}

The atom mirror is placed facing upwards in a vacuum chamber
pumped to $5\times10^{-9}$ Torr.  In contrast to the experiments
that use a standard magneto-optical trap (MOT) to trap and cool
the atoms $\sim2$ cm above the mirror, we use the mirror MOT
technique to collect the atoms 1.5 to 4 mm above the
surface~\cite{Reichel}.  A MOT requires the zero of a magnetic
quadrupole field to be centered at the intersection of six
circularly polarized laser beams coming from all cardinal
directions.  To satisfy this configuration near the hard drive
surface, two 1 cm diameter beams of opposite circular polarization
reflect at 45$^\circ$ from the 1 cm$^2$ etched region (see
Fig.~\ref{fig:mMOT}).  A retroreflected beam is positioned
perpendicular to the 45$^\circ$ beams and grazes the surface of
the hard drive.  Aligning the axis of the quadrupole field with
one of the 45$^\circ$ beams completes the mirror MOT
configuration.  The trapping lasers, each with an intensity of 4
mW/cm$^2$ and 1 cm wide, are detuned by 10 MHz from cesium's
$F=4$, $F'=5$ cycling transition. A repumping beam tuned to the
$F=3$, $F'=4$ transition is superimposed onto both the grazing
beam and a 45$^\circ$ beam. The atoms are loaded from a thermal
vapor.

In previous experiments using a perfectly reflecting gold mirror, we have been
able to trap $2\times10^6$ cesium atoms in a mirror MOT and cool them to 3
$\mu$K. One might expect trapping and cooling to be much less effective with
the etched hard drive due to its poor qualities as an optical mirror: the
reflectivity is only $\sim50$\%, it is a good optical grating, and the
magneto-optical Kerr effect degrades the circularity of the reflected
45$^\circ$ beams. Nevertheless, we have been able to collect $1\times10^6$
atoms and sub-doppler cool them to 11 $\mu$K. Achieving this low temperature
is crucial because the atoms released directly from the mirror MOT, at a
temperature of $\sim120$ $\mu$K, expand too quickly and become too diffuse to
detect by the time they reach the hard drive surface.

The poor optical reflectivity of the mirror does slightly complicate the
sub-doppler cooling procedure; however with careful zeroing of the magnetic
field it is still possible to achieve polarization-gradient cooling to 11
$\mu$K in a (downwards) moving reference frame. The atoms are optically pumped
into the $F=4$, $m_F=4$ Zeeman sub-state just before being dropped, and we
apply a 100 mG bias field parallel to the magnetic stripes in order to
maintain alignment of the atomic spins while they are falling/bouncing.

\begin{figure}
\scalebox{.5}[.5]{\includegraphics*[.8in,2.6in][7.6in,8.15in]{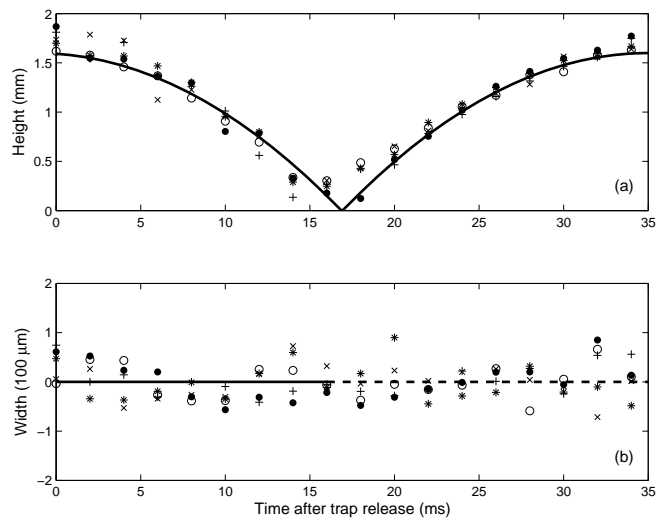}}
\caption{\label{fig:data} Panel (a) shows the mean height of the
atoms above the hard drive surface during the first bounce.  Panel
(b) shows the residuals from a linear fit to the width of the
expanding atom cloud (see text).}
\end{figure}

We have been able to detect two full bounces of the atoms from the
hard drive atom mirror. Figures~\ref{fig:data}(a) and (b) show
data from five runs of the experiment. The top panel shows the
mean position of the atoms above the hard drive surface as a
function of time. Superimposed is a curve depicting the expected
trajectory of a particle falling under gravity and bouncing from a
hard wall. The slope of a line fit to the lateral expansion of the
falling atom cloud provides a measure of the atoms' rms velocity.
A non-specular mirror would heat and diffusely scatter the
reflected atoms as they bounce, resulting in a sharp increase of
the cloud expansion rate.  We made a linear fit to pre-reflection
($t<15$ ms) data in each of the data sets, and deviation from this
line, post-reflection, would be evidence of non-specularly. The
dashed segment demarcates the region of unfitted data, and we do
not see any increase or offset of the residuals in this
post-reflection region: to within the experimental resolution, we
do not detect any deviation from specular reflection.

We have realized a specular atom mirror built by etching a common hard drive.
Magnetization periodicity of 3 $\mu$m has been achieved, and we believe it
would be straightforward to reduce this to 2 $\mu$m with photolithography and
to $\sim1$ $\mu$m using a large area electron-beam writer. The hard drive atom
mirror is compact, passive, relatively simple to fabricate, and possesses a
large remanent magnetic field. Moreover, it has several desirable properties
for applications beyond simple reflection of atoms. The hard drive's large
coercivity should allow one to use wires fabricated directly on its surface to
augment the mirror's ability to manipulate atoms. Likewise, electric pads
could be printed on the surface. These pads would allow state-independent
forces to act in concert with the state-dependent forces from the mirror's
magnetic field to perform quantum logic gates necessary for quantum
computation~\cite{Tommaso}. The mirror can trap cold atom gases in 2D, and can
act as an adjustable grating when used in conjunction with a magnetic bias
field~\cite{Hinds00b,Hinds00a}. Large area mirrors can be fabricated, and it
seems possible that these mirrors could be useful for guiding or confining
cold neutrons~\cite{Vlad}. As hard drive platters are expected to have good
surface flatness and substrate rigidity, it may be possible to create 2D
waveguides and other devices by holding an opposing pair of atom mirrors a few
microns apart.

\begin{acknowledgments}
The authors thank the Roukes group and Marko Loncar. This work was
supported by the Multidisciplinary University Research Initiative
program under Grant No. DAAD19-00-1-0374.
\end{acknowledgments}

\bibliography{ambib}

\end{document}